\documentclass{emulateapj}
\usepackage{ulem}  
\usepackage{color} 
\newcommand{\rel}{relativistic} 
\newcommand{\nonrel}{non-relativistic} 
\newcommand{\tSNR}{t_\mathrm{SNR}} 

\newcommand{\effDSA}{\epsilon_\mathrm{DSA}}

\newcommand{\RXJSNR}{SNR RX J1713.7-3946}

\shorttitle{Thermal Emisssion Spectra from Cosmic Ray Modified Shocks}

\begin{document}

\title{The Role of Diffusive Shock Acceleration on NonEquilibrium
Ionization in Supernova Remnant Shocks II: Emitted Spectra}

\author{Daniel~J. Patnaude \altaffilmark{1},
 Patrick Slane\altaffilmark{1},
\& John~C. Raymond\altaffilmark{1}, Donald~C. Ellison\altaffilmark{2}}

\altaffiltext{1}{Smithsonian Astrophysical Observatory, Cambridge, MA 02138}

\altaffiltext{2}{Physics Department, NC State University, Box 8202,
Raleigh, NC 27695; don\_ellison@ncsu.edu}

\begin{abstract}

  We present a grid of
  nonequilibrium ionization models for the X-ray spectra from supernova
  remnants undergoing efficient diffusive shock acceleration.  The
  calculation follows the hydrodynamics of the blast wave as well as the
  time-dependent ionization of the plasma behind the shock. The
  ionization state is passed to a plasma emissivity code to compute the
  thermal X-ray emission, which is combined with the emission from
  nonthermal synchrotron emission to produce a self-consistent model for
  the thermal and nonthermal emission from cosmic-ray dominated
  shocks. We show how plasma diagnostics such as the $G'$--ratio of
  He-like ions, defined as the ratio of the sum of the intercombination,
  forbidden, and satellite lines to the resonance line, can vary with
  acceleration efficiency, and discuss how the thermal X-ray emission,
  when the time-dependent ionization is not calculated self-consistently
  with the hydrodynamics, can differ from the thermal X-ray emission
  from models which do account for the hydrodynamics. Finally we compare
  the thermal X-ray emission from models which show moderate
  acceleration ($\sim$ 35\%) to the thermal X-ray emission from
  test-particle models.

\end{abstract}


\section{Introduction}

In young supernova remnant (SNR) shocks, the acceleration of cosmic rays
(CRs) leads to a softening of the equation of state in the shocked 
plasma. This comes about because the diffusive shock acceleration (DSA)
process turns some \nonrel\ particles into \rel\ ones and because the
highest energy \rel\ particles escape from the shock. Both of these effects
lead to lower post-shock plasma temperatures as well as higher 
post-shock densities \citep[e.g.,][]{jones91,berezhko99}. The ionization
of shocked gas at a particular time depends upon both the gas
density and the electron temperature, and it has been recently shown
by \citet{patnaude09} that the time-dependent
nonequilibrium 
ionization (NEI) and electron temperature, $T_e$, are influenced by the DSA 
efficiency, $\effDSA$.

A number of young SNRs show some evidence for 
both nonthermal and thermal emission behind
the forward shock, including SN~1006 \citep{vink03,bamba08}, Tycho
\citep{hwang02}, Kepler \citep{reynolds07}, and recently
Cas A \citep{araya10}. The thermal emission arises when the forward 
shock sweeps through the circumstellar medium (CSM) and heats it to 
X-ray emitting temperatures. As pointed out in \citet{ellison07}, the thermal
X-ray emission is often considerably fainter than the nonthermal emission,
but there exist examples where the thermal emission is as bright or brighter
than any nonthermal emission, such as in parts of RCW~86 \citep{vink06} and
N132D \citep{xiao08}. In contrast, in \RXJSNR, the lack of thermal X-ray emission
is an important constraint on the ambient medium density, and we
have recently shown that if the TeV $\gamma$-ray emission is hadronic in 
origin, then copius thermal X-ray emission would be observed 
with current X-ray observatories \citep{ellison10}.

Here we extend our previous work \citep{patnaude09,ellison10} 
by coupling our models to a plasma emissivity code,
producing thermal X-ray spectra and ionization timescale parameters
({\it i.e.} $n_e t$) as a function of position and time in the SNR
shock. Again, as in \citet{patnaude09}, 
we limit our analysis to the region between
the forward shock (FS) and contact discontinuity (CD). In \S~2 we
present our new model and outline how we have coupled our hydrodynamical
model to a plasma emissivity code. The inclusion of the hydrodynamical
evolution makes it difficult to compare the computed thermal emission to
existing models, since the shocked, ionizing plasma is expanding
adiabatically and the electron temperature is constantly evolving.  To
test the code, we present in \S~2 a test where we follow the ionization
in the case of a fixed density and temperature, which is equivalent to
available NEI models in spectral fitting codes such as {\sc Xspec}.  

The
parameter space involved in producing the thermal and nonthermal
emission as well as the hydrodynamical evolution is extensive and in
\S~3 we show a simplified set of models where we only vary a subset of
parameters such as the shock acceleration efficiency. In this section,
we show how the so called $G$-ratio, which is a measure of
electron temperature, varies with acceleration efficiency. We find
that since the increased acceleration efficiency generally leads to
lower electron temperatures, increasing the acceleration efficiency
results in higher $G$-ratios. In \S~4, we compare our models to similar
models where the X-ray emission is calculated not from the ionization
state of each element, but from the ionization timescale  
which the shocked gas is expected to have at the end of the simulation.
We find that the results differ, mainly at lower energies. Finally in
\S~4 we discuss the recent result by \citet{berezhko10} and show that
the absence of thermal X-ray emission does not naturally lead to the
interpretation that the TeV gamma-ray emission is hadronic in origin. 
We argue that if the blastwave has only recently hit the circumstellar
shell, then the shocked gas, though at X-ray emitting temperatures,
will be underionized. We show that our results and those of 
\citet{ellison10} are consistent with the material being at a lower 
density. 

\section{Emissivity Code}

In previous papers \citep{ellison07,patnaude09,ellison10}, 
we outlined our model which coupled the NEI calculation to the cosmic
  ray acceleration and hydrodynamics code. For each simulation cell and
  timestep, we calculate an ionization state vector $f(X^{i})$, an
  electron temperature $T_e$, and electron and ion densities, $n_e$ and
  $n_i$. Combined with a chosen set of abundances (here we choose those
  of \citet{anders89}), these quantities are all that is required as
  input to produce the thermal X-ray emission. As in \citet{patnaude09},
the ionization processes included in our NEI calculation include
direct collisional ionization and collisional excitation followed by
autoionization. Additionally, the recombination rates include 
radiative recombination and dielectronic recombination. We neglect
Auger ionization as well as photoionization. In the current version
of our code, we do not include ionization from super-thermal
particles; in an ionizing plasma,
these effects are expected to be small \citep{porquet01a}.

We have chosen to couple our models to an updated version of the 
Raymond--Smith emissivity code \citep[RS93;][]{raymond77,brickhouse95}. 
RS93 differs from other available
codes in terms of the number of included emission lines, a few line centroids,
and individual line emissivities. These differences are sufficiently 
numerous to make it difficult to compare our calculated spectra to other
available models such as APED, but the differences are well 
documented \citep[c.f., ][]{smith01}. Additionally, since our models not 
only follow the time dependent ionization, but also are dynamical
in nature, it is not adequate to compare the calculated spectra to existing
models such as \texttt{nei} and \texttt{pshock} in {\sc Xspec}. 

\begin{figure}
\epsscale{0.5}
\plotone{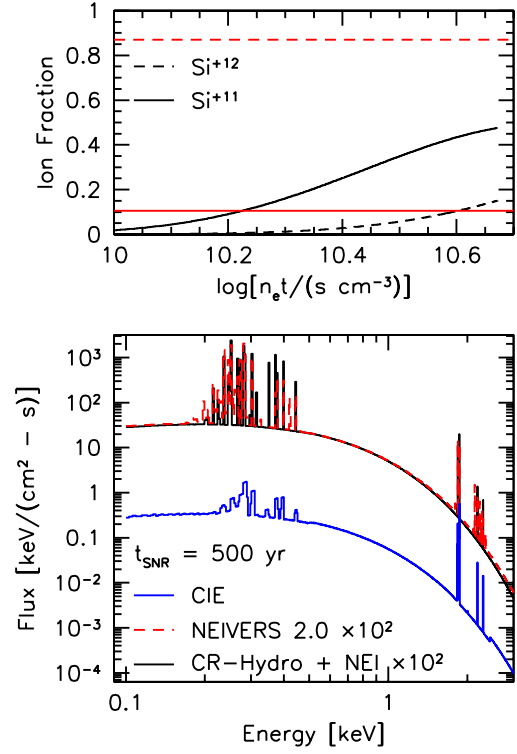}
\caption{{\it Top Panel}: Evolution of ionization fraction 
for Si$^{+11}$ and Si$^{+12}$ in a single temperature nonequilibrium
ionization calculation up to $\tSNR$ = 500 yr. The solid and dashed red 
lines correspond to the collisional ionization equilibrium values for
Si$^{+11}$ and Si$^{+12}$ at $T_e$ = 3 $\times$ 10$^{6}$ K.
{\it Bottom Panel}: Thermal X-ray emission
line spectrum for a single temperature NEI model at an ionization 
age of 4.7$\times$10$^{10}$ s cm$^{-3}$, 
which corresponds to $\tSNR$ = 500 yr and
n$_{p,0}$ = 3.0 cm$^{-3}$. We only include emission from
H, He, and Si. The black curve corresponds to our calculation
and is compared directly to an equivalent {\sc Xspec} \texttt{nei}
model at the same ionization age and temperature, which
is shown as the red curve. The blue, curve
corresponds to the emission from a plasma at the same temperature
in collisional ionization equilibrium.}
\label{fig:test_t500}
\end{figure}

To determine whether the modeled spectra are consistent with existing
NEI calculations, as well as consistent with models in collisional
ionization equilibrium, we performed the following test: we ran a
simulation for a prescribed amount of time, and 
far upstream from the shock, we fix the electron
temperature at a constant value of $T_e$ = 3$\times$10$^{6}$ K and keep
the density fixed at $n_{p,0}$ = 3.0 cm$^{-3}$. That is, while in general
the hydrodynamics computes the electron temperature and densities 
self-consistently behind the shock, for the purposes of this test we
keep the electron temperature at a high, constant value far enough
upstream of the shock so that we can evolve the NEI. 
This is equivalent to the single temperature, single
ionization age models found in {\sc Xspec}, such as \texttt{nei} and
\texttt{vnei}. We compare our computed spectrum to both what is computed
in the {\sc Xspec} model as well as a model for a plasma in collisional
ionization equilibrium (e.g. \texttt{raysmith} or \texttt{mekal}). For
simplicity, we limited our calculations to spectra that only include H,
He, and Si. In Figure~\ref{fig:test_t500}, we show the computed and
reference spectra for a shocked plasma that has been evolving for 500
yrs, as well as what the same spectrum looks like under conditions of
collisional ionization equilibrium. In the top panel of
Figure~\ref{fig:test_t500}, we also plot the evolution of Li- and
He-like ions of silicon.

At an age of 500 years, corresponding to an ionization age of
$\sim$ 5$\times$10$^{10}$ s cm$^{-3}$, at an assumed density of 
$n_{p,0}$ = 3.0 cm$^{-3}$, we observe emission lines
from several atomic transitions. For instance, from the helium iso-sequence, 
we see
line emission at 2.295 keV (1s$^2$ - 1s4p), 2.185 keV (1s$^2$ - 1s3p)
1.866 keV (1s$^2$ - 1s2p), and the multiplet transitions 
1s$^2$ - 1s2p$^3$P$^0$ at 1.85 keV and 1s$^2$ - 1s2s$^3$S. Also,
at lower energies we see numerous emission lines from the lithium to 
nitrogen iso-sequences, including multiplet transitions which lead
to multiple spectral lines.

\begin{figure}
\epsscale{0.5}
\plotone{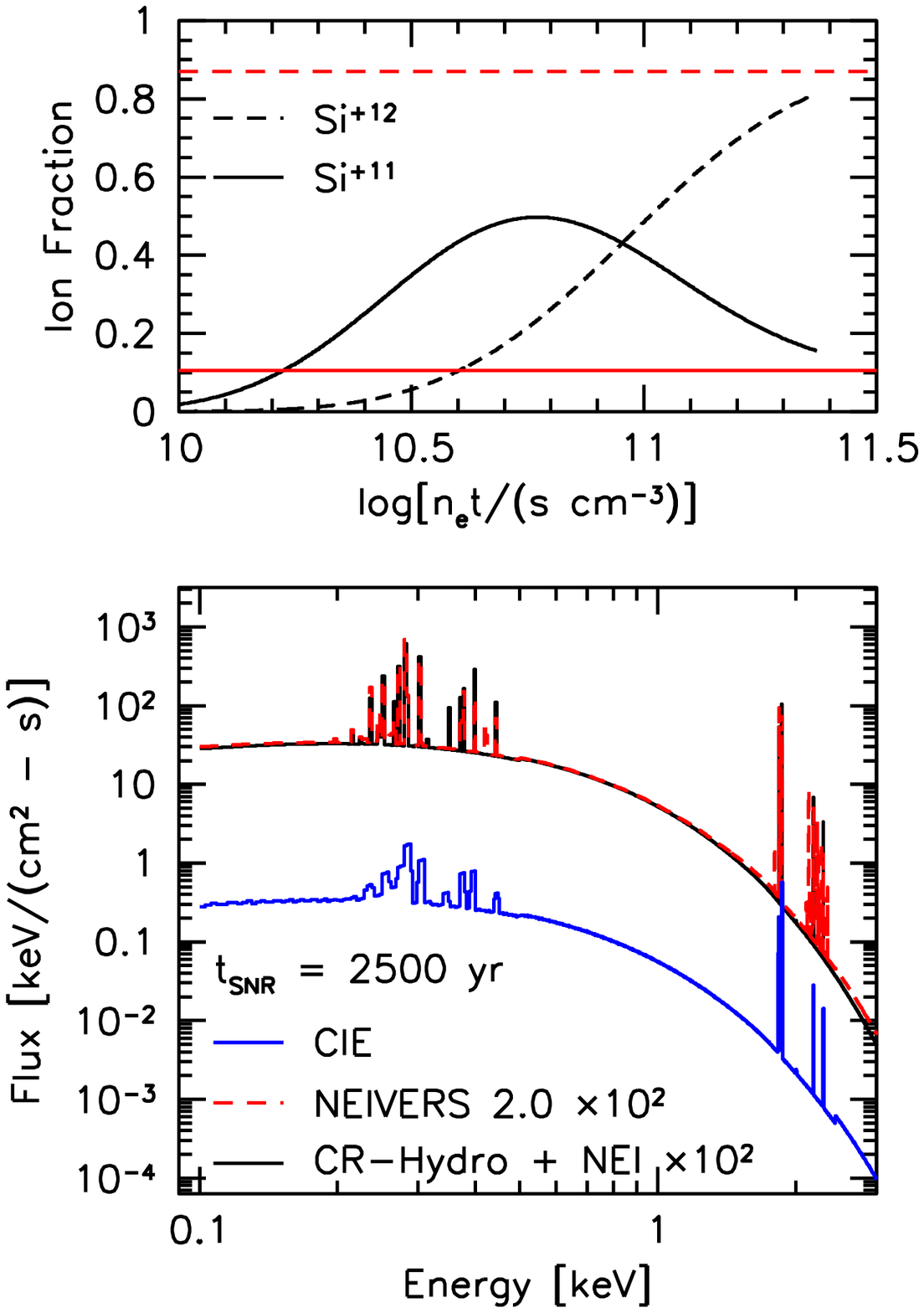}
\caption{{\it Top Panel}: Evolution of ionization fraction for 
Si$^{+11}$ and Si$^{+12}$ in a single temperature nonequilibrium
ionization calculation up to $\tSNR$ = 2500 yr. The solid and dashed red 
lines correspond to the collisional ionization equilibrium values for
Si$^{+11}$ and Si$^{+12}$ at $T_e$ = 3 $\times$ 10$^{6}$ K.
{\it Bottom Panel}: Thermal X-ray emission
line spectrum for a single temperature NEI model at an ionization 
age of 2.4$\times$10$^{11}$ s cm$^{-3}$, 
which corresponds to $\tSNR$ = 2500 yr and
n$_{p,0}$ = 3.0 cm$^{-3}$. We only include emission from
H, He, and Si. The black curve corresponds to our calculation
and is compared directly to an equivalent {\sc Xspec} \texttt{nei}
model at the same ionization age and temperature, shown
here as a red curve. The blue, curve
corresponds to the emission from a plasma at the same temperature
in collisional ionization equilibrium.}
\label{fig:test_t2500}
\end{figure}

In Figure~\ref{fig:test_t2500}, we plot the same model 
($T_e$ = 3$\times$10$^{6}$ K and $n_{p,0}$ = 3.0 cm$^{-3}$) 
at an age of 2500 yrs. After 2500 years, the same helium-like lines are
seen as in the $\tSNR$ = 500 year model, but the intensities have increased
relative to the lower ionization states. This is reflected in the
top panel of Figure~\ref{fig:test_t2500}, where the lithium-like
state of silicon is approaching the collisional ionization equilibrium
value of $\sim$ 0.1 (at $T_e$ 3$\times$10$^{6}$ K.) Likewise, in collisional
ionization equilibrium at 3$\times$10$^6$ K, the He-like state is
populated at $\sim$ 0.87 (in fact, in collisional ionization 
equilibrium, Li-- and He-like states are the
dominant species at this temperature for silicon, with additional 
contributions from the Be-like state; a negligible fraction of
ions are in the H-like state), and the NEI conditions
shown here are seen to be approaching it after 2500 years.

 
In both test cases, the computed and reference NEI spectra are in good
agreement, showing the same underlying continuum and similar line emission, 
with differences most noticeable in the emission lines between
200 and 500 eV. We attribute these differences to small differences between
the underlying atomic data used in the codes. The most obvious difference
between the two is seen as the obvious lack of an emission line in the
{\sc Xspec} \texttt{nei} model at $\sim$
350 eV (Si XII n=2 to n=4 multiplet). 
In the $\tSNR$ = 500 year model, this line is quite bright, while in the higher
aged model it appears to be weakening, relative to the surrounding lines
(which are also weaker, relative to the continuum. 

The goal of this testing was to ensure that the NEI calculation in our model, 
in the limit of a single ionization age, single temperature \texttt{nei} model,
is consistent with existing models. Realizing that there are 
differences in the underlying atomic data that will be manifest in the 
existence
and intensity of individual emission lines (in particular, the 
\texttt{nei} model includes many more emission lines from He-like states,
as seen in Figures~\ref{fig:test_t500} and ~\ref{fig:test_t2500}), 
our aim was to track the trending in the
two NEI models to ensure that they behave consistently against one another. As 
an additional test, we chose to compare our spectrum to that from a plasma in
collisional ionization equilibrium. We find that, as expected, the computed 
NEI is in fact approaching the CIE spectrum expected for the temperature and
density we are using for the test \citep{smith10}. 

\section{Results}

The cosmic ray hydrodynamics code contains an extensive set of parameters
which affect the SNR dynamics, emitted thermal and nonthermal broadband
spectrum, and relativistic particle populations. An exhaustive study of
the effects that all these parameters have on the emitted spectrum is
beyond the scope of this paper, but we note that combinations of parameters
have been successfully used to produce emitted spectra from \RXJSNR, 
ruling out certain models for that SNR \citep{ellison10}. To simplify 
the process, we choose a set of models where we only vary the acceleration
efficiency $\effDSA$ and ambient medium density, and leave everything else fixed. 
We note that the 
application of these models to a specific SNR would require a very specific
set of parameters, so we restrict ourselves to phenomenological studies for
now. 

Our model assumes the following set of parameters: $\tSNR$ = 
1000 yr, E$_{\mathrm{SN}}$ = 10$^{51}$ erg, and M$_{ej}$ = 1.4M$_{\odot}$. 
Additionally, we assume an exponential ejecta
profile and an ambient magnetic field of 15$\mu G$. Parameters which 
affect the ionization calculations and resultant thermal emission include
the assumed abundances and the electron heating. We assume cosmic 
abundances \citep{anders89} and heating via Coulomb collisions. The ambient medium
has a temperature of 10$^{4}$ K and is preionized at 10\%. We vary the
acceleration efficiency, $\effDSA$ between 1\% (test particle) and
75\%, and except where indicated otherwise, we chose ambient medium
densities, $n_{p,0}$ = 0.1, 0.3, 1.0, and 3.0 cm$^{-3}$. Our model
also includes the ability to apply interstellar absorption as well as
combine the thermal and nonthermal emission. For clarity, we do not include
any absorbing column, and except where noted, we neglect the contribution to the spectrum
by the nonthermal continuum. We note, as in \RXJSNR, that the underlying
nonthermal continuum can be the dominant source of X-ray photons, so 
the models like those presented here are most appropriate for comparison
with emission line fluxes for shocks that
show a combination of thermal and nonthermal emission.

The emitted thermal spectrum is a combination of bremsstrahlung 
and two photon continua and
line radiation. The formation of lines mainly occurs by electron 
impact of atoms or ions.  The excitation
can generally be broken down into excitation of outer-shell 
electrons, excitation of inner-shell electrons, and resonant excitation
\citep{raymond96}.  
Additional contributions from radiative recombination
to excited states, dielectronic recombination satellite lines and
inner shell ionization are sometimes significant, especially in
plasmas out of ionization equilibrium.  In the process of radiative 
recombination, the following occurs with

\begin{equation}
Z^{+(z+1)} + e^{-} \rightarrow Z^{+z}(n) + h\nu_{cont} \rightarrow Z^{+z}(m) + h\nu_{mn} \, ,
\end{equation}

\noindent
where the ion $Z^{+(z+1)}$ captures an electron and is de-excited
to the $m$-th excited state via emission of a photon.
In the process of dielectronic recombination, a
free electron gains kinetic energy as it approaches an ion, then
excites a bound electron to an energy level higher than 
the energy it had at infinity, so the free electron is 
captured to form a doubly excited state of
ion $Z^{+z}$,

\begin{equation}
Z^{+(z+1)} + e^- \rightarrow (Z^{+z})^{**} \, .
\end{equation} 

\noindent
If auto-ionization occurs (the reverse of the capture process) the system
returns to its original state and no recombination takes place. 
Alternatively a fraction of the ions in the autoionizing state decays
by spontaneous radiative transition of the inner excited electron 
to a state below the first ionization limit:

\begin{equation}
(Z^{+z})^{**} \rightarrow (Z^{+z})^{*} + h\nu \, ,
\end{equation}

\noindent
where the stabilizing transition in the recombined ion $Z^{+z}$ results in
the emission of a dielectronic satellite line of the parent transition in the
recombining ion $Z^{+(z+1)}$. Eventually, the singly excited state
cascades down to the ground state with subsequent emission of photons.

\subsection{Emission Line Diagnostics}

To demonstrate a potentially useful diagnostic predicted by the 
models that should
be observable by {\it Astro-H}, 
we consider the n=2 $\rightarrow$ n=1 emission of
He-like ions.  This consists of a resonance ({\it r}) $2^1P \rightarrow 1^1S$
line, a forbidden ({\it f}) $2^3S \rightarrow 1^1S$ line, and an
intercombination ({\it i}) $2^3P\rightarrow 1^1S$ line. The intensity
ratio (f+i)/r, called the $G$--ratio, varies with density, ultraviolet
radiation intensity, temperature and ionization state. The shocked, swept-up
material behind SNR shocks is considered to be at a low enough density
($n_e << 10^{10}~\rm cm^{-3}$), and the UV radiation
field is sufficiently negligible, that density dependent variations in the
spectra are not observable. Therefore we will consider only the variation with
temperature and ionization state. As a point of reference, the
$G$--ratio decreases with increasing electron temperature \citep{gabriel69}. Finally, we note
that there are a number of satellite lines ({\it S}) formed by transitions of
the form $\rm 1s2s^2 \rightarrow 1s^22s,
1s2s2p \rightarrow 1s^22p ~and~ 1s2p^2 \rightarrow 1s^22p$  which lie at
wavelengths near the i and f transitions. Since current and near term observations
are unable to resolve their contributions to the spectrum, we define an 
observable ratio $G'$ = (f+i+s)/r.

The lines in the $G'$--ratio contain contributions from direct collisional
excitation by electrons ($\it f, i$ and $\it r$), radiative and dielectronic
recombination of the H-like ion into the excited levels ($\it f, i$ and $\it r$),
dielectronic recombination of the He-like ion ($\it s$), innershell excitation
of Li-like and lower ions ($\it s$) and innershell ionization of Be-like
and lower ions ($\it s$).

Since the resultant emitted thermal X-ray spectrum can be quite complicated,
containing emission lines from H- and He-like states as well as a forest
of emission lines from intermediate charge states (e.g. as was 
demonstrated
in Figures~\ref{fig:test_t500} and ~\ref{fig:test_t2500}), 
we focus on the bulk characteristics of the emitted spectrum. To do
this, we sum the emitted spectrum from the contact discontinuity to 
forward shock and investigate the global characteristics of the 
thermal emission as a function of both acceleration efficiency, and to 
a lesser extent the ambient medium density.

\begin{figure}
\epsscale{0.5}
\plotone{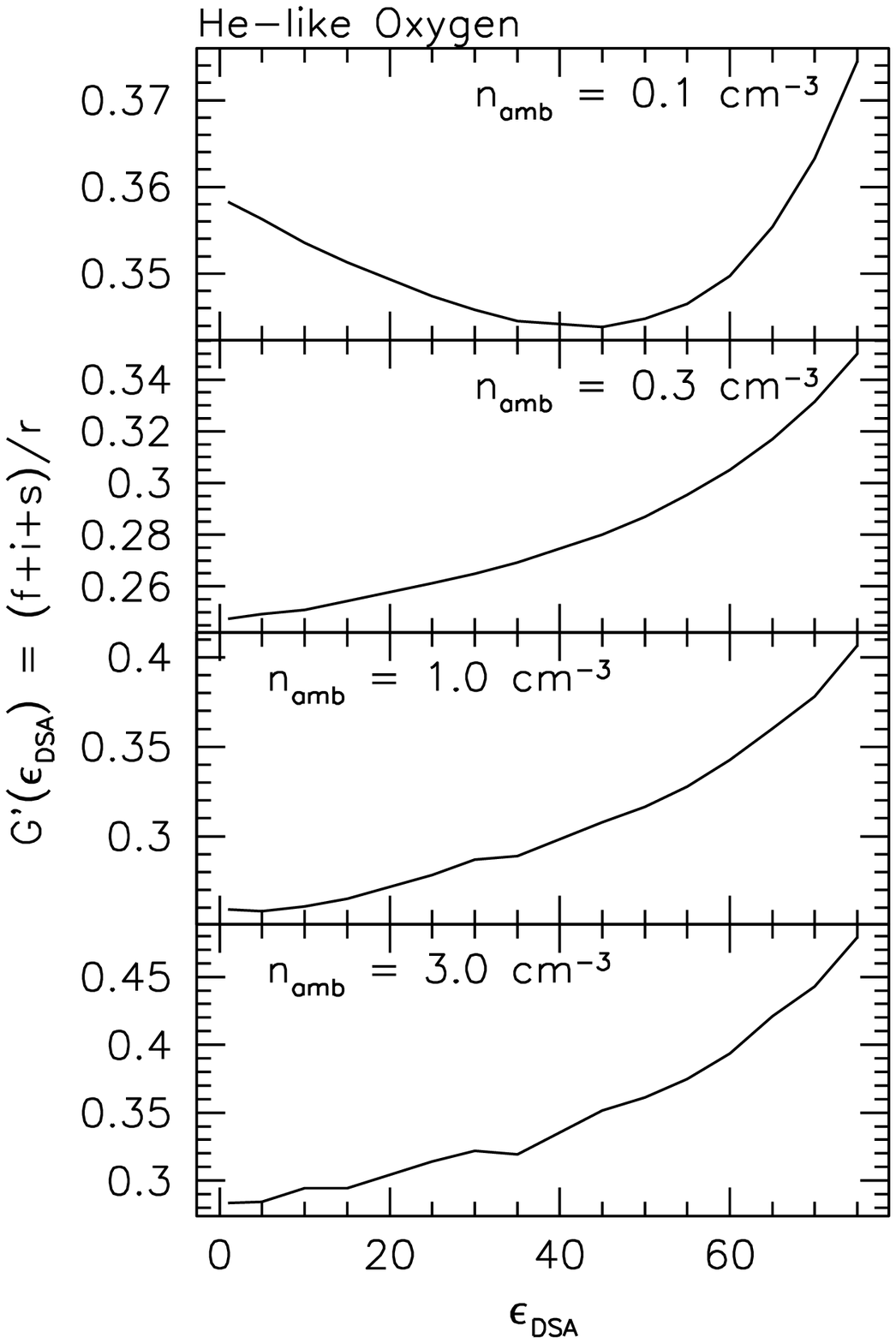}
\caption{$G'$--ratio for He-like Oxygen in models with 
$n_{p,0}$ = $0.1$, $0.3$, $1.0$, and $3.0$ cm$^{-3}$ versus acceleration
efficiency. The $G$--ratio is integrated over the entire region between
the forward shock and contact discontinuity.}
\label{fig:o_g_ratio}
\end{figure}

\begin{figure}
\epsscale{0.5}
\plotone{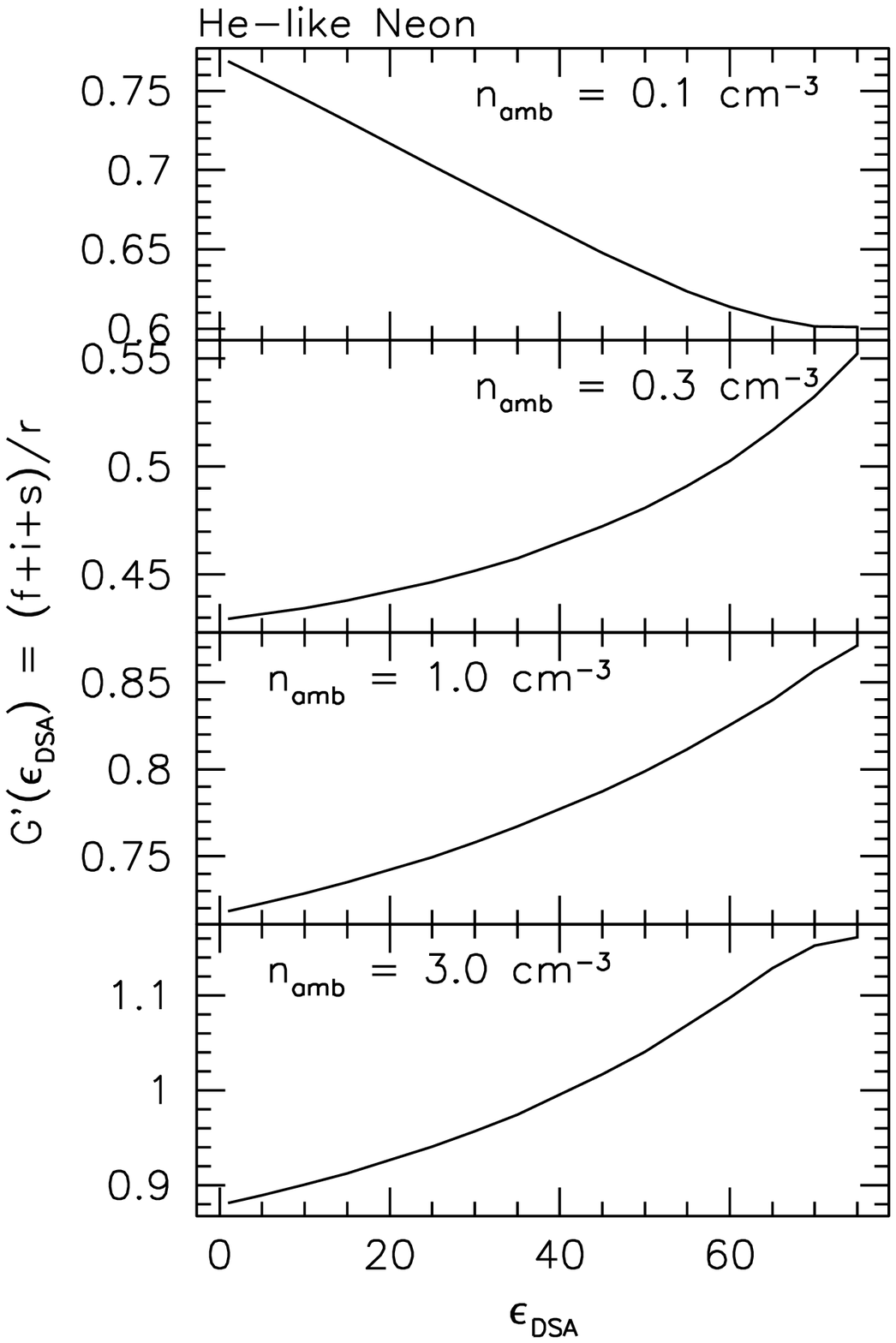}
\caption{$G'$--ratio for He-like Neon in models with 
$n_{p,0}$ = $0.1$, $0.3$, $1.0$, and $3.0$ cm$^{-3}$ versus acceleration
efficiency. The $G$--ratio is integrated over the entire region between
the forward shock and contact discontinuity.}
\label{fig:ne_g_ratio}
\end{figure}

\begin{figure}
\epsscale{0.5}
\plotone{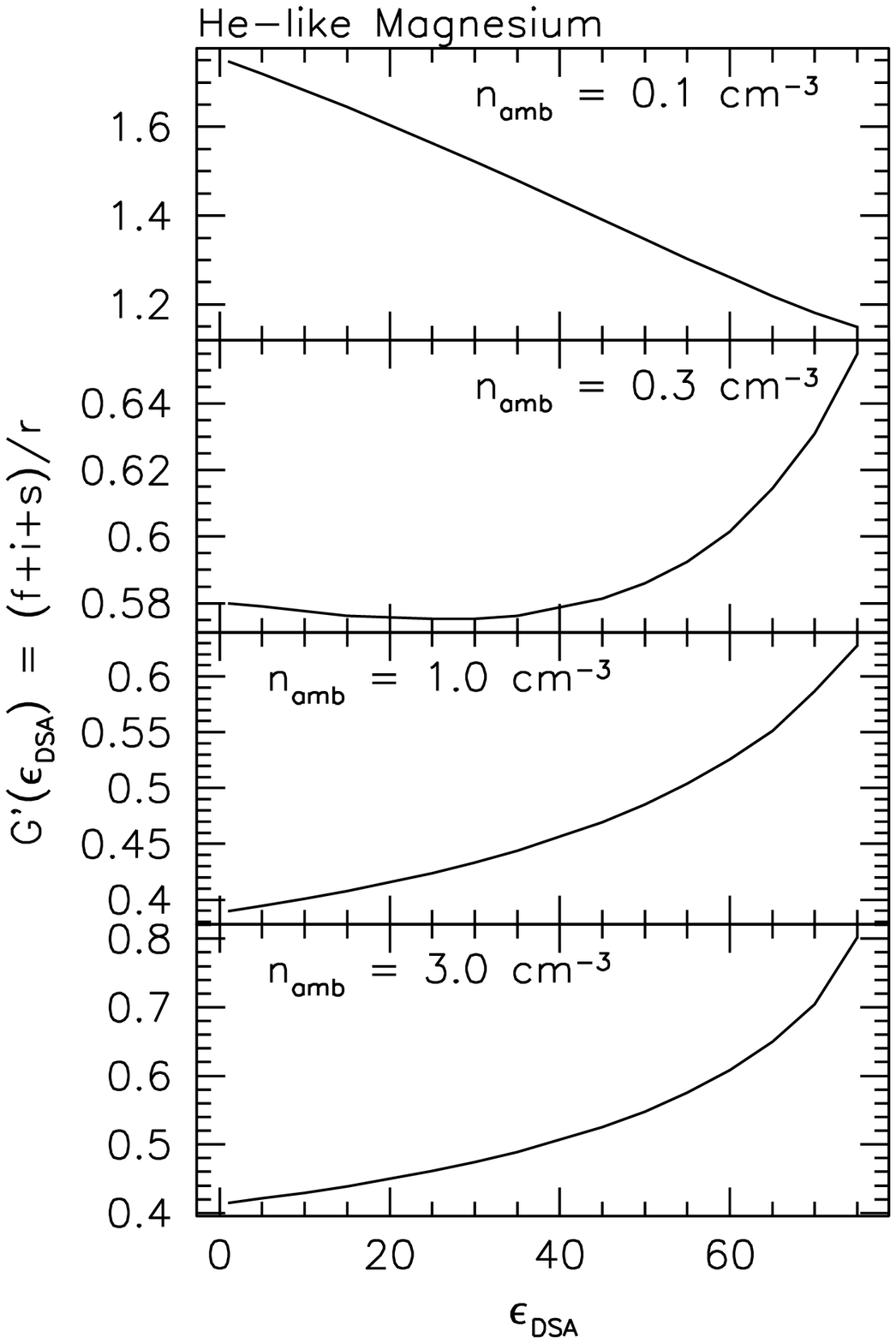}
\caption{$G'$--ratio for He-like Magnesium in models with 
$n_{p,0}$ = $0.1$, $0.3$, $1.0$, and $3.0$ cm$^{-3}$ versus acceleration
efficiency. The $G$--ratio is integrated over the entire region between
the forward shock and contact discontinuity.}
\label{fig:mg_g_ratio}
\end{figure}

\begin{figure}
\epsscale{0.5}
\plotone{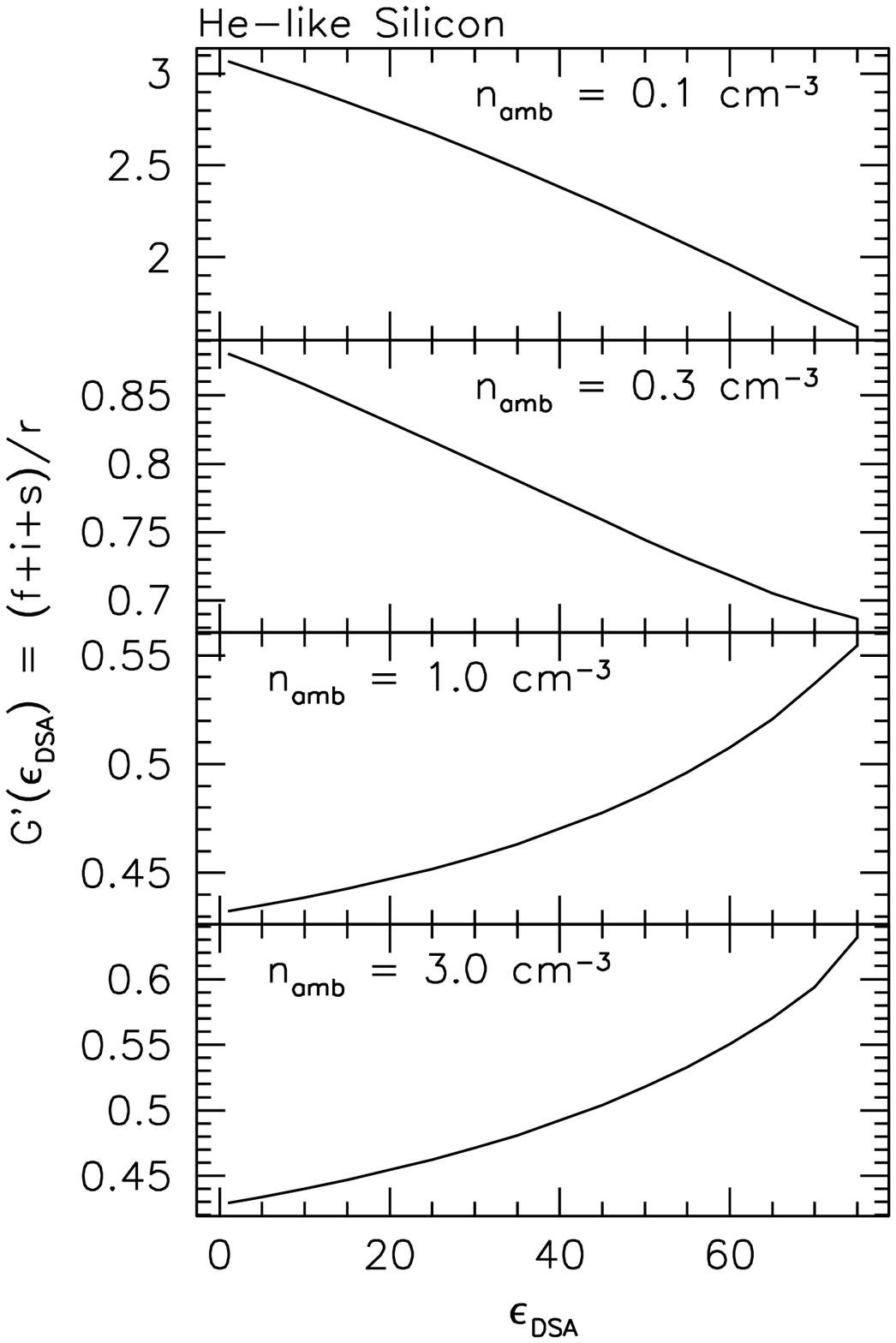}
\caption{$G'$--ratio for He-like Silicon in models with 
$n_{p,0}$ = $0.1$, $0.3$, $1.0$, and $3.0$ cm$^{-3}$ versus acceleration
efficiency. The $G$--ratio is integrated over the entire region between
the forward shock and contact discontinuity.}
\label{fig:si_g_ratio}
\end{figure}

In Figures~\ref{fig:o_g_ratio} -- ~\ref{fig:si_g_ratio} we plot the $G'$--ratio
for the He-like states of oxygen, neon, magnesium, and silicon. Because 
excitation cross sections for the forbidden and intercombination lines decrease
with energy, relative to that of the resonance line, and because the cross sections
for the forbidden and intercombination lines have strong resonance contributions
near the excitation threshold, $G'$ can be thought of as a diagnostic of the
electron temperature. In addition, the dielectronic recombination satellites are excited
by electrons well below the threshold for exciting the $\it r$ line, but since
they are observed near the triplet state (where they remain unresolved with 
current instrumentations), their contribution to the flux should be included, thus
weakening the overall dependence upon temperature.
The $G'$--ratio also depends upon the ionization state of the gas, since the triplets
are populated by recombination from the H-like states, and innershell
excitation and ionization of lower charge states also contribute. The interplay
of temperature and ionization state requires a fully time-dependent ionization model such
as we present here, and
this ratio will be an important diagnostic tool for upcoming missions such as {\it Astro-H}.
We note that we have so far neglected excitation and ionization by the nonthermal
electrons (e.g. \cite{porquet01a}), and that some improvements to
the atomic rates in our code remain to be implemented (e.g., \cite{porquet01b}),
but that we expect the trends of the predictions to be reliable.

In \citet{patnaude09} we showed that the electron temperature varied roughly inversely
with the acceleration efficiency, that is, as acceleration efficiency increases,
the electron temperature decreases. Therefore, in order to understand how the
$G'$--ratio changes, we plot it versus the acceleration efficiency, rather than 
electron temperature.
In Figure~\ref{fig:o_g_ratio}, we show the variations in $G'$ of oxygen with acceleration
efficiency, at constant density. At high ambient densities $G'$ behaves as expected. 
That is, $G'$ increases with increasing efficiency (or, equivalently, increases with
decreasing temperature). However, the 
lowest density ambient medium model shows the opposite trend, except at the highest 
acceleration efficiencies, where $G'$ is seen to rise rapidly. Similarly, in neon, as 
seen in Figure~\ref{fig:ne_g_ratio}, the
$G'$--ratio decreases with increasing efficiency for $n_{amb}$ = 0.1 cm$^{-3}$.
However, at a density of 0.3 cm$^{-3}$, the trend with acceleration efficiency
is consistent with the high density results for oxygen. A similar result is seen for
magnesium, while in silicon, the variation in $G'$ at $n_{amb}$ = 0.1 and 0.3 cm$^{-3}$
is opposite to that seen in the higher density models (compare the upper two
panels of Figure~\ref{fig:si_g_ratio} to the bottom two). 
The decline in $G'$
with acceleration efficiency in the lower density models can be understood as
an effect of the underionization of the plasma.  An increase in acceleration efficiency
leads to a higher compression ratio, and since the ionization rates of ions below
He-like are not very sensitive to temperature for these ions at the temperatures
of these models, the higher acceleration efficiency models reach higher He-like ionization
fractions. This increases the strength of the resonance line relative to the inner shell 
excitation lines.

\subsection{Global Properties}

\begin{figure}
\epsscale{0.5}
\plotone{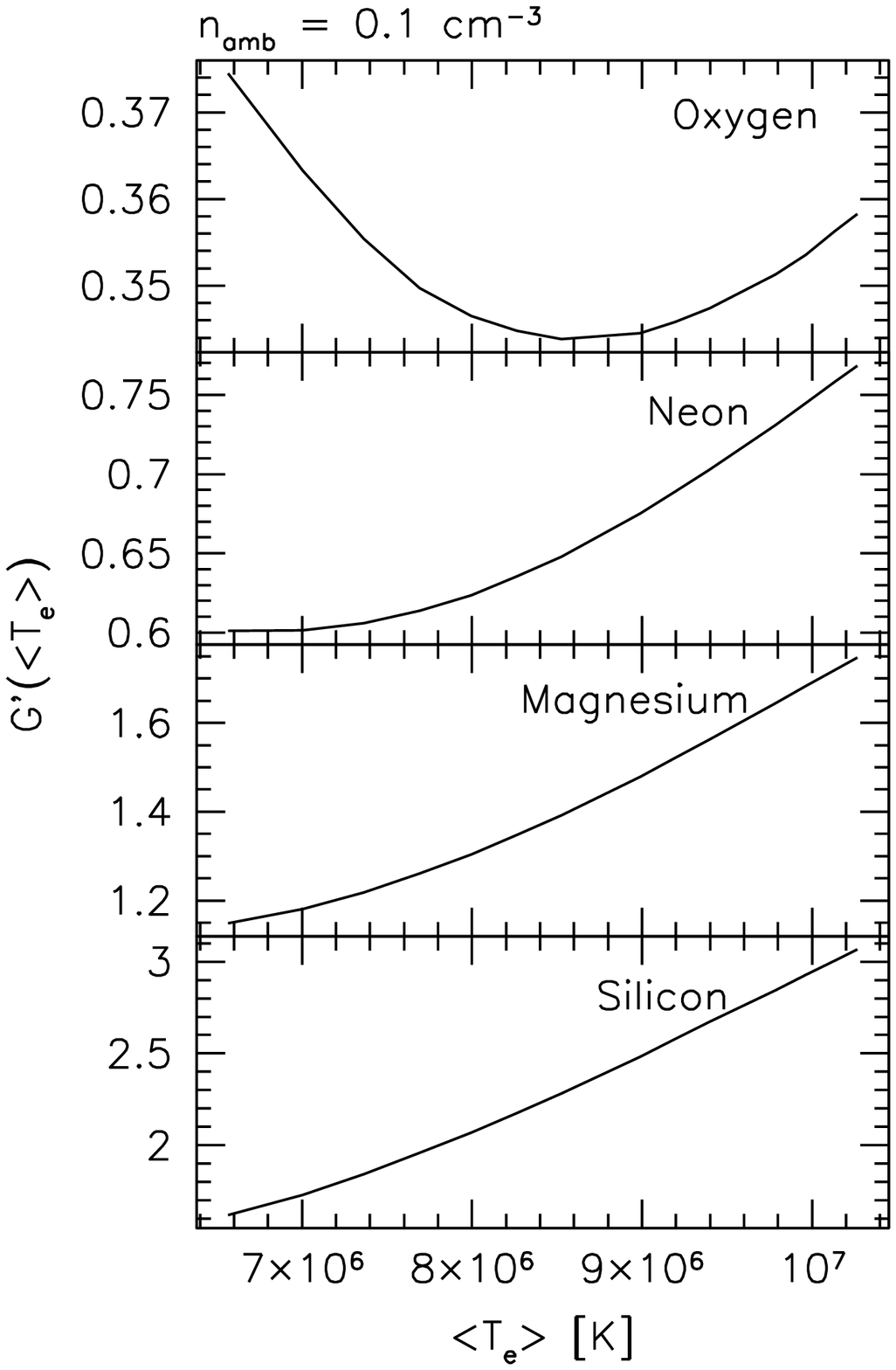}
\caption{$G^\prime$--ratio as a function of emission measure weighted electron
temperature for oxygen, neon, magnesium, and silicon, for an ambient
medium density of 0.1 cm$^{-3}$.}
\label{fig:te_p1cc}
\end{figure}

\begin{figure}
\epsscale{0.5}
\plotone{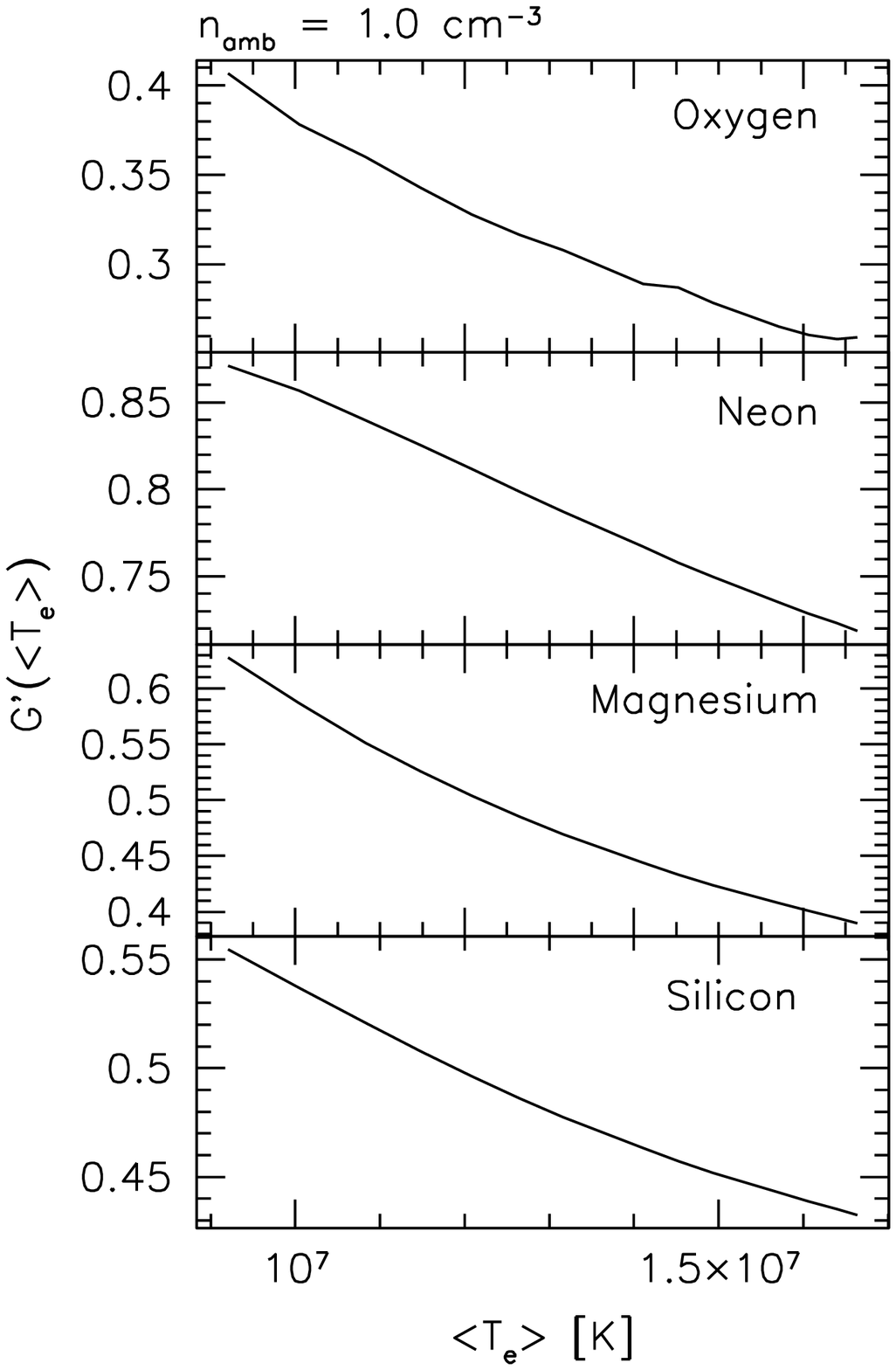}
\caption{$G^\prime$--ratio as a function of emission measure weighted electron
temperature for oxygen, neon, magnesium, and silicon, for an ambient
medium density of 1.0 cm$^{-3}$.}
\label{fig:te_1cc}
\end{figure}

To better understand the variations seen in $G'$ with $\effDSA$, we plot in 
Figures~\ref{fig:te_p1cc}--\ref{fig:te_1cc} the variation in $G'$ as a function
of weighted electron temperature, at constant ambient medium density. The
weighted electron temperature is over the entire region between the forward
shock and contact discontinuity, and we weight the
electron temperature in each hydrodynamical cell by the emission measure of
the cell (i.e. $<T_e>$ = EM$^{-1}$ $\Sigma_i (T_e^i \times EM^i)$). The
emission measure may not be the best quantity to weight the 
temperature by, and it may be more meaningful to weight the temperature by the
power in the He-like triplet, but that will result in a weighted temperature which
varies with element, such that the average temperature as derived from the
He-like oxygen will be different from the average temperature as derived
by silicon. Using the emission measure to weight the temperature allows
us to compare how $G'$ varies with $T_e$ for each element, and also compare
how it varies between elements. 
We note that the emission measure in each cell is the only quantity
related to the emitted spectrum which will not vary with element. 

In Figure~\ref{fig:te_p1cc} we plot $G'$ versus $T_e$ for O, Ne, Mg, and Si, at
constant ambient medium density. The emission measure weighted 
average electron temperature for the shocked CSM varies from $\sim$ 
7$\times$10$^{6}$ K ($\sim$ 600eV) in the highly efficient models to $>$ 10$^7$
K ($\sim$ 850 eV) in the test particle models. The curves are complicated;
in the top panel, O is seen to decrease initially with increasing $<T_e>$ before
increasing above 9$\times$10$^6$ K. However, everywhere the resonance line
intensity is well above the sum of the forbidden, intercombination, and 
satellite lines ($G'$ $<$ 1.). In neon, again the resonance line is brighter
than the other contributions, while in magnesium and silicon, the
flux from the forbidden, intercombination, and satellite lines is greater
than for the resonance line, and increases with increasing temperature
(i.e. decreasing acceleration efficiency).

Generally, the $G'$--ratio is sensitive to temperature; at high temperatures,
all states are collisionally excited, but at lower temperatures, dielectronic recombination
becomes important. The triplet states will be preferentially populated through
recombinations simply because of their greater statistical weight, thus 
increasing the intercombination and forbidden line strengths relative to the
resonance line. This is seen in the low temperature (high acceleration efficiency)
end of Figure~\ref{fig:te_1cc}, where the $G'$--ratio is highest. This is not,
however, the case for the lower density model. Here, the $G'$--ratio is generally seen
to be highest in the test particle model.

\begin{figure}
\epsscale{0.5}
\plotone{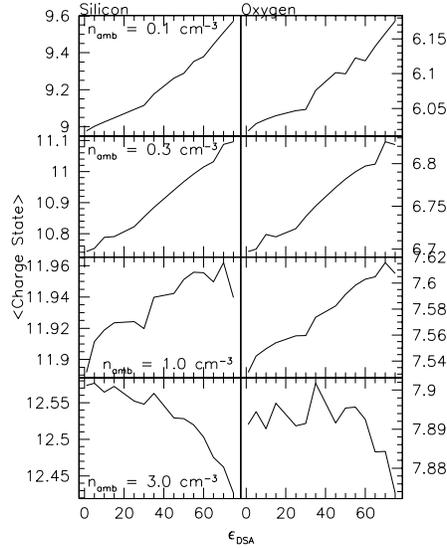}
\caption{Average integrated charge states for silicon (left hand panels)
and oxygen (right hand panels) as a function of density and acceleration
efficiency. The higher density models show some scatter but, like the
lower density models, there appears to be a clear trend with acceleration
efficiency.}
\label{fig:charge}
\end{figure}

To understand the differences between Figure~\ref{fig:te_p1cc} and ~\ref{fig:te_1cc},
we plot in Figure~\ref{fig:charge} the average charge state for silicon and
oxygen. Two trends are immediately 
clear from this plot. First, in the low density
case, the average charge state for silicon is well below the He-like state, but
the average charge state increases with acceleration efficiency. Additionally,
a higher ambient medium density results in a higher
average charge.  This is because the higher density behind the shock
increases the collisional ionization rate, and this dominates over the
temperature dependence of the ionization rates of Si XII and lower ions. 
At the highest densities, the average charge is much higher, and it
{\it decreases} with increasing efficiency.  In this case, the decline 
in temperature with acceleration efficiency decreases the ionization rate
of Si XIII, and this dominates over the increase in density because
of the high ionization potential of He-like Si XIII. As seen in 
Figure~\ref{fig:charge} similar trends are seen in oxygen. Here, however,
even at the highest efficiencies in the high density cases, 
oxygen is almost fully ionized.

Equating these results to Figures~\ref{fig:te_p1cc} and ~\ref{fig:te_1cc} means
that in the low density case (Fig.~\ref{fig:te_p1cc}), the lower density 
implies less collisional ionization and less efficient Coulomb heating, so that
the plasma stays less ionized. More precisely, in the low density models,
the He-like states are not reached, and forbidden line emission is dominated
by inner shell excitation of lower ions. In higher density models, the He-like
states are reached, and the expected dependence with temperature
is seen (c.f.~Fig.~\ref{fig:te_1cc}). We stress that in the low density models,
some elements, such as silicon, will be highly underionized and the He-like
triplets will be quite faint. These lines will be hard to detect, and when
combined with any nonthermal continuum emission, they might be undetectable
with current instrumentation.

\begin{figure}
\epsscale{0.5}
\plotone{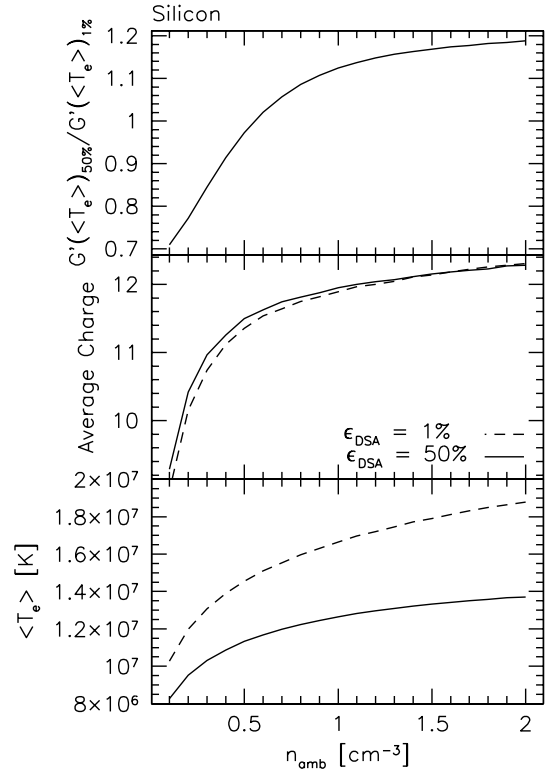}
\caption{{\it Top}: The ratio of $G^\prime$ for He-like silion 
in a 50\% efficient model versus a 
test-particle model for a range of densities between 0.1 -- 2.0 cm$^{-3}$. 
{\it Middle}: The average charge state behind the shock for the test-particle
and efficient model. {\it Bottom}: The emission measure weighted electron
temperature.}
\label{fig:consteff}
\end{figure}

We end this section with Figure~\ref{fig:consteff}, 
a plot of the variation in $G'$, average charge state,
and electron temperature as a function of ambient medium density, at constant
acceleration efficiencies. This plot shows how these values vary between
the test particle case and a 50\% efficient shock. As expected, the average
electron temperature in the test particle case is higher than in the
efficient case, but the average charge states are very similar. 
The biggest differences come from the ratio of $G'$, where in the
low density models, the $G'$--ratio is higher in the test particle case
than in the efficient case, while above a density of $\la$ 1.0 cm$^{-3}$, the
relation is reversed, with the $G'$--ratio in the efficient case is higher. 
From this plot, it is clear that at identical ambient medium densities, 
similar charge states can result between the efficient and inefficient models,
while the temperatures can differ by $\sim$ 30\% and the line ratios can
differ by $\sim$ 10\%. For instance, for an ambient medium density of 1.0 cm$^{-1}$,
the weighted temperature in the 50\% efficient model is 30\% lower than 
in the test particle case. However, the ionization state for silicon here
is virtually identical, but the $G'$--ratio in the efficient model is 10\%
higher than in the inefficient model. At temperatures which can differ by
as much as they do in Figure~\ref{fig:consteff}, what we see here is that
at constant ambient medium density but differing efficiencies, different lines
are seen at different intensites. Since the line ratios and underlying 
electron temperatures are different between the efficient and inefficient 
models, the shape of the resultant spectra will differ.

\section{Discussion}

The models presented here differ from existing available models that
compute the X-ray emission from shocks in several ways, but the relevant
difference here is that the nonequilibrium ionization is followed 
simultaneously with the shock dynamics and particle acceleration. The 
ionization vector is then passed directly to an emissivity code to compute
the thermal X-ray emission. Since the NEI is evolved simultaneously with
the hydrodynamic evolution, our calculation includes changes to the NEI
as a result of adiabatic losses, or as is the case in an efficient model,
increased compression behind the shock resulting from efficient shock 
acceleration. 

\begin{figure}
\epsscale{0.5}
\plotone{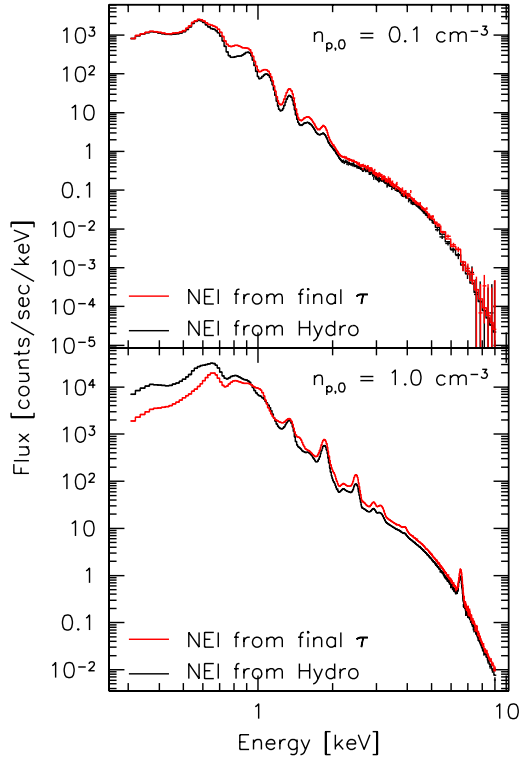}
\caption{{\it Top}: Simulated {\it Chandra} ACIS-S X-ray spectrum from a
model with n$_{p,0}$ = 0.1 cm$^{-3}$ where the ionization is calculated
self-consistently with the hydrodynamics (black curve) versus the ionization
being calculated after the simulation, based on the final ionization age
and temperature (red curve). {\it Bottom}: Same as in the top panel, but for
an ambient medium density of n$_{p,0}$ = 1.0 cm$^{-3}$. In both cases,
we assume $\effDSA$ = 1.0\%.}
\label{fig:comp}
\end{figure}

To illustrate this effect,
in Figure~\ref{fig:comp}, we compare the thermal X-ray emission from
models where we follow the time-dependent ionization with the hydrodynamics
(black curves) to the thermal X-ray emission from models where
we calculate the ionization structure from the final ionization age ($\tau$)
and electron temperature (red curves). Here, the final ionization 
for each cell age is defined as the $\tau$ at the end of the simulation.
Then, for comparison, we re-run the NEI calculation for the
final electron temperature and the time equivalent to
the age of the SNR at the end of the hydrodynamical
simulation. For these models, we do not
include any efficient particle acceleration, as that will only complicate
the interpretation of these results. Computing the ionization structure
in each grid cell from the final $\tau$ is similar to assuming a 
single $\tau$, single $T_e$ model for each cell in {\sc Xspec} (e.g.
the \texttt{nei} model), except that in our computation of $\tau$
and $T_e$ in the hydrodynamics, we still account for adiabatic cooling
in the shocked gas, since these values were computed initially
from the hydrodynamics. This approach is similar to \citet{ellison07}. As seen 
in Figure~\ref{fig:comp}, there are differences between the two
resultant spectra. In the low density (n$_{p,0}$ = 0.1 cm$^{-3}$) model,
the calculation of the emitted spectrum from the final ionization
age slightly overpredicts the final thermal X-ray emission, most
evidently seen around the He-like lines of magnesium and silicon, as
well as from the emission of L-shell iron. Not surprisingly, the
shape of the underlying continua are consistent. This is expected
because the shape of the continuum is determined by the electron
temperature, which are identical between the two models.

In the higher density (n$_{p,0}$ = 1.0 cm$^{-3}$) model, the differences
are much larger, particularly at low energy. In particular, the ionization state 
and thermal X-ray emission computed from the final ionization age and
temperature underpredict the Fe-L emission. We interpret this
difference as being a direct result of not following the ionization
explicitly; that is, when computing the ionization state from a
single density and $\Delta t$, intermediary ionization states are
not correctly populated, since the calculation assumes that
the shocked gas has gone from a cold (10$^{4}$ K) gas to hot (several
10$^{6-7}$ K) in a single timestep. 
We note that were a spectrum like this fit with existing
models, it would appear that the shocked plasma is overabundant
in metals such as argon and calcium. Figure~\ref{fig:comp} clearly
demonstrates the need to compute the ionization self-consistently
with the shock hydrodynamics (regardless of whether one considers
the effects of efficient diffusive shock acceleration).

\subsection{Efficient vs. Inefficient Models}

\begin{figure}
\epsscale{0.5}
\plotone{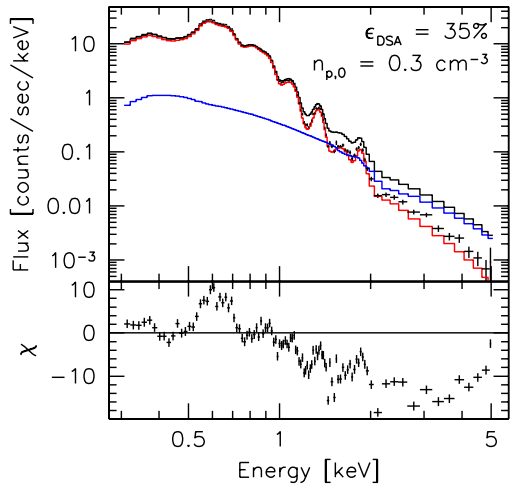}
\caption{Comparison between the emitted thermal X-ray spectrum from a
test-particle model versus the combined thermal and nonthermal 
spectrum from a 35\% efficient model, with n$_{p,0}$ = 0.3 cm$^{-3}$. The
data correspond to the simulated test-particle spectrum, the black curve
corresponds to the combined thermal and nonthermal spectrum from the
efficient model, the blue curve corresponds to the synchrotron component
contribution to the model, and the red curve corresponds to the thermal
contribution. The bottom panel shows the difference between the
thermal emission from the test particle model (the data) and the
model (here, the thermal emission from the efficient model as well as the
contribution from the continuum.)}
\label{fig:tp_vs_eff}
\end{figure}

In Figure~\ref{fig:tp_vs_eff}, we plot the observed differences between
the thermal emission from a test-particle model and the combined
thermal and nonthermal emission from an efficient model.
The black crosses in Figure~\ref{fig:tp_vs_eff} are simulated data for
a 50 ksec Chandra ACIS-S observation of an SNR with $\effDSA$
= 0 and $n_{p,0}$ = 0.3 cm$^{-3}$. The blue curve is the nonthermal
emission, the red curve is the thermal emission and the
black curve is their sum. In the efficient model, we assume
$\effDSA$ = 35\% and $n_{p,0}$ = 0.3 cm$^{-3}$. 
As expected, at the
high energy end, the nonthermal emission dominates the shape of the 
spectrum. However, below about 2 keV, where the the thermal
emission is comparable to the nonthermal emission, differences are still seen. 

Below 1 keV, the thermal emission is 
$\sim$ an order of magnitude above the nonthermal emission. It is in
this regime where the differences between the thermal emission from
an efficient model and a test partcle model are most apparent. In
particular, the emission from oxygen is seen to differ between these
two models, and is reflected in the fit residual, shown in the
lower panel, which shows that the modeled thermal emission from the efficient
case does not accurately describe the thermal emission in the test particle
case. This suggests that the emission lines contribute at differing
levels between the efficient and inefficient models.

\subsection{Comparisons to Recent Calculations for Limits on the
Thermal X-ray Emission in RX J1713.7-3946}

Recently, \citet{berezhko10} used models from \citet{hamilton83}
to constrain the origin of the TeV $\gamma$-ray emission in
RX J1713.7-3946. They argued that, based on fits to the broadband nonthermal
continuum, the TeV $\gamma$-ray emission is hadronic in 
origin, with $\sim$ 35\% of the shock kinetic energy being deposited
into nuclear cosmic rays . However, \citet{ellison10} argue that if
the TeV $\gamma$-ray emission were hadronic in origin, thermal 
X-ray emission would be detected, though the model used there is for
a constant ambient medium density of 0.2 cm$^{-3}$, for the hadronic
scenario. In the model developed by \citet{berezhko10}, 
the SNR blastwave has been 
expanding through a low density ($n_H$ $\approx$ $0.0008$ cm$^{-3}$)
bubble and has recently encountered the bubble wall, with
$n_H$ = 0.25 cm$^{-3}$ at present. From their model, we
estimate that the shock began encountering the bubble wall $\sim$ 800 years
ago, as determined by the curve for $N_g$ in the top panel of their 
Figure~1 \citep{berezhko10}. They assume that the thermal
X-ray emission as determined for the Sedov solution by 
\citet{hamilton83} is reduced by an amount equivalent to 
the ratio of the bubble emission measure to the Sedov emission 
measure, here equal to 0.46 \citep{berezhko10}. Additionally, they assume
a value of $\eta$ = 8$\times$10$^{49}$, where $\eta$ = n$^2$E$_{\mathrm{SN}}$
\citep{hamilton83}. Their chosen value of $\eta$ corresponds to an 
explosion energy of 1.3$\times$10$^{51}$ erg and an ambient medium 
density of 0.25 cm$^{-3}$. They use the value of $\eta$ and an
assumed electron temperature of 1 keV to estimate the
thermal X-ray emission from the recently shocked bubble wall, and
they find that at 1 keV, the thermal X-ray emission
is approximately half the nonthermal flux \citep{berezhko10}. 

Since the blastwave hit the bubble wall $\sim$ 800 years ago
and the electron temperature is $\sim$ 1 keV, the shocked plasma
will be underionized. This is probably best illustrated in Figure~\ref{fig:consteff},
where for an ambient medium density of 0.25 cm$^{-3}$, the average charge
state of silicon is quite low, and in fact since the model shown in 
Figure~\ref{fig:consteff} is for a constant density model, it represents
an upper limit on what the ionization state would be in \citet{berezhko10}. Thus,
the shocked plasma from the bubble wall will likely be underionized.


Basically, \citet{berezhko10} use a model for
thermal X-ray emission from a Sedov-like SNR (of age 1600 yr and
distance 1 kpc, but renormalized to match the emission measure of a
shocked shell) to the emission from a recently shocked thin
shell undergoing efficient particle acceleration. 
As we showed above, however, The emitted spectra in the two scenarios 
are expected to be markedly different. We stress
that in order to constrain the origin of the TeV gamma-ray emission, whether
it arises from either leptonic or hadronic sources, the time-dependent
ionization and thermal X-ray emission must be calculated in a self-consistent
manner with the shock acceleration and hydrodynamics, as was done
for a simple case in \citet{ellison10}. 

As was shown in previous sections,
the increased compression from the shock acceleration increases the
line intensities (from oxygen and neon, for instance), so \citet{berezhko10}
will underestimate the contributions of line emission to the total
flux. This is consistent with \citet{ellison10} where both a low ambient medium density
and a high electron-to-proton ratio at the maximum energies
were required in order to hide any emission lines and still fit the
shape of the GeV--TeV continuum.

Since the shock acceleration does has an effect on the ionization balance and thus the
emitted thermal X-ray spectrum, using thermal X-ray emission models which 
do not account for increased ionization or compression when trying to understand
the origin of GeV--TeV emission is not accurate. This will be of 
particular
importance when the blastwave is moving through complex environments such
as circumstellar shells, where the density can change by a few orders of magnitude
over relatively small spatial scales. In \RXJSNR, the absence of thermal X-ray
emission rules out a hadronic origin to the $\gamma$-ray emission in the case
of expansion into a uniform circumstellar medium \citep{ellison10}. Additional modeling will be
required to assess whether or not the hadronic origin to the TeV emission is a
viable picture under other, more complex conditions.

\section{Conclusions}

We have presented an extensive grid of models where we study the
effects on the emitted thermal X-ray spectrum behind SNR shocks. To
accomplish this, we
have coupled our CR-Hydro+NEI calculation to an updated version
of the Raymond--Smith plasma emissivity code. In studying the
global properties of the shocked plasma, we find that in the
low density limit, the plasma temperature diagnostic, the $G'$--ratio,
decreases with decreasing temperature (increasing acceleration
efficiency) for higher Z elements such as magnesium and silicon. 
This is opposite to the behaviour at higher densities, where 
the $G'$--ratio increases with decreasing temperature, as expected.
We find that the reason for this behavior is that in the
lower density models, the electron temperatures are lower, but
more importantly, the ions do not reach the
higher ionization states. Additionally, we find that in low density 
models, the
average charge states increase with increasing efficiency, while
at higher densities, the charge state is seen to decrease. This is
because in the low density model, the higher efficiency leads to 
higher post shock densities and therefore more collisional excitation,
while in the high density models, the shocked
plasma does not reach as high a temperature in the efficiently
accelerating models, so the ions are less ionized, even though the post
shock density is also high. The differences in average charge
state between the low and high density models are thus viewed as a result
of the temperatures in the low density model generally being higher
than in the high density model.

Based on our results, we conclude that it is not sufficient to calculate
the ionization state of the plasma as a ``post-processing'' step. While
in lower density models the differences are small, at higher
densities, the calculation will underpredict the emission from intermediate
charge states, particularly emission from Fe-L. We speculate that these
differences might be even larger in metal rich ejecta where the
number of free electrons will be large. Finally, we note that
the shape of the emitted spectrum in an efficient model, even when
folded through a CCD-resolution response, is fundamentally different
than the emitted spectrum from a test-particle model. 

\acknowledgements

D.~J.~P. acknowledges support from {\it Chandra} Theory grant
TM0-11006A, and P.~O.~S. and D.~J.~P. acknowledge support from NASA
contract NAS8-03060. D.C.E. acknowledges support from NASA contracts
ATP02-0042-0006, NNH04Zss001N-LTSA, and 06-ATP06-21.  The authors are
grateful to the KITP in Santa Barbara where part of this work was done
when the authors were participating in a KITP program.

\end{document}